# Effect of Lubricant Contaminants on Tribological Characteristics During Boundary Lubrication Reciprocating Sliding

Mohamed Kamal Ahmed Ali [*], Fawzy M.H. Ezzat, K.A.Abd El-Gawwad, M.M.M.Salem.

Automotive and Tractors Engineering Department, Faculty of Engineering, Minia University, 61111, El-Minia, Egypt.

[*] *Corresponding author: Eng.m.kamal@mu.edu.eg (M. K. A. Ali)*

**Abstract:**

This article presents the effect of the presence of solid contaminants, in engine lubricating oil, on the tribological parameters, which impact negatively on the performance of the engine and increase fuel consumption. This study revealed that the lubricant was contaminated by Fe, Cu, Al, Pb and $SiO_2$ particles. The tribological tests were performed using 0.63, 0.85 and 1.1 m s$^{-1}$ average sliding speeds and 120 N contact load to mimic the boundary lubrication regime of the sliding reciprocating motion of the piston ring/liner interface in an engine. The presence of the solid contaminants in engine oils leads to an increase in friction coefficient, wear and frictional power losses, increasing the surfaces roughness as a consequence. The results showed that grain size and concentrations strongly affected the tribological parameters. In order to minimize the effect of solid contaminants, it is necessary to improve the filtration accuracy for lubricating oils.

**Keywords:** Solid contaminants, Engine oils, Boundary lubrication, Friction, Wear.

## 1. Introduction

The main purpose of lubricating oils is the reduction of friction between surfaces, prevention of wear and rust, cooling by removing the heat resulting from the contact of the surfaces and the cleaning of automotive engines as a protection from damage caused by friction and wear. Mechanical frictional losses in automotive engines vary between 17% and 19% of the total energy generated by an engine.[1, 2] The piston ring-cylinder liner contributes approximately 40% to 50% of frictional losses in automotive engines. The sliding contact between engine parts comprises a variety of different friction and wear mechanisms during one working cycle of the engine. Due to the variations in speed, load and counter surface effects, the lubrication conditions in an engine are strongly transient, which is reflected by variations in the friction and wear behavior. Maximum values of friction coefficient at top dead center and bottom dead center locations were in the range of 0.10 to 0.15 with mid-stroke values ranging from 0.05 to 0.10 for the study conducted by Ali.[3] These values depend on the particular lubricant used, the surface quality and surface material.[4]

The major sources of these contaminants in lubricating oils are prior blowby, lubricant breakdown, and wear of engine parts. Solid contaminants such as iron (Fe), copper (Cu), aluminum (Al), lead (Pb) and silicon oxide ($SiO_2$) are a result of the wear of the engine parts as a consequence of external contaminants mixing with the lubricating oil, fuel and intake air. Figure 1 displays the types of contamination in pneumatic, fluid, and solid forms as well as its effects. The damage of the parts is dependent on grain size of particles, concentrations, texture of the particles and operating pressure.[5] The solid contaminants mainly consist of particles.[6] The presence of solid contaminants enhances the oxidation of lubricating oils.[7] The lubrication regimes for the piston ring assembly depend on contact load, the kinematic viscosity of lubricating oils, sliding speed and surface roughness.[8] A high level contaminant concentration in the form of solid particles leads to high thin-film wear at the start of sliding.[9] The presence solid contaminants in lubricating oils responsible for the failure of machine parts.[10]

The reason for an engine oil change could be due to its deterioration in terms of viscosity and oxidation, as well as solid contaminants that become mixed or dissolved in the lubricating oils. The presence of contaminants in engine oil is generally undesired, as solid contaminant particles are a potentially cause of abrasive wear. On the other hand, liquid contaminants may cause corrosive, viscosity and tribochemical wear changes.[11, 12] Contamination of lubricating oils causes wear of piston ring, which generates more contamination. This proceeds via internal wear generating fresh wear debris leading to the opening of the dynamic sealing surfaces. Bore polishing is another undesired impact of small abrasive particles the lubricating oils, while larger particles can cause scratches in the bore.[13] The total friction in an engine immediately after a cold-start is four to five times higher than at fully warmed-up conditions.[14] The source of the contamination particles may be soot from combustion, silica dust and similar minerals, and wear particles consisting of ferrous, lead, chromium, copper, aluminum, nickel alloys and tin[15] The diesel soot interacts with lubricating oils and ultimately leads to wear of engine parts.[16, 17]

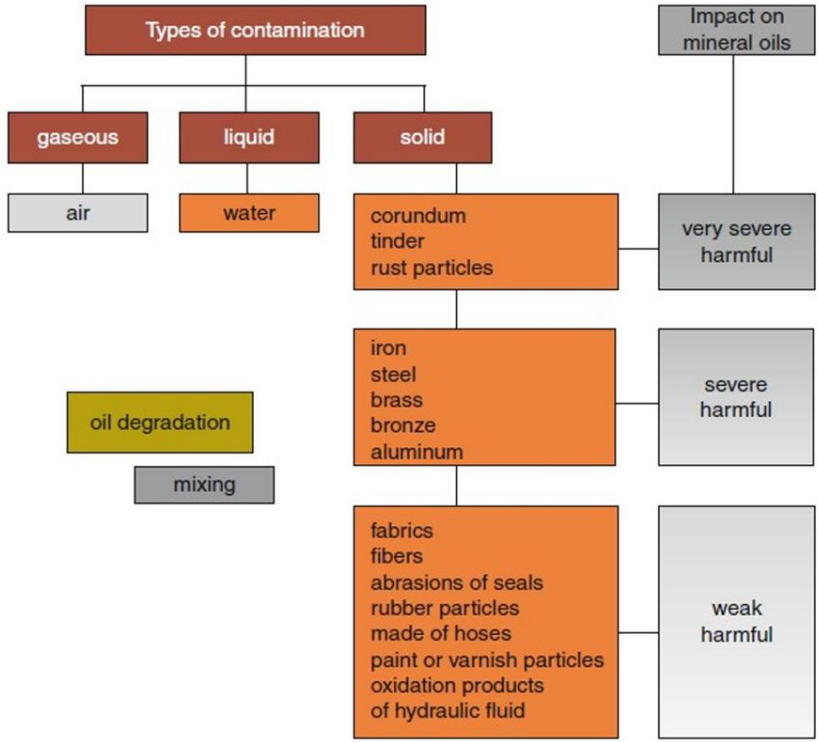

**Figure 1.** Contamination types in lubricating oils and Harmful effects.[4]

The harmful effect of solid particles in the lubricating oils is particularly obvious with softer parts like piston skirts, bearings, and cam-follower contacts.[18] The major categories of solid contaminant particles in crankcase oils are carbon, or combustion particles and metallic or wear particles. The carbon contaminants particle can be determined by means of Fourier transform infra-red spectroscopy (FTIR), a method that is suitable for the determination of the presence of different organic compounds including reaction products.[19] It is necessary for all lubricated systems that the particle size should remain well below the oil film thickness between surfaces in any lubricated mechanism, and this very well applies to engines. Oil filters in experimental applications can have nominal retention rates in the order of 15 to 20 µm. Lubricant oil in the region of the piston rings is far more contaminated than the oil in the engine sump. Exhaust gas recirculation without soot filters cause an increase in the level of carbon particles in the lubricating oil.[20] An increase in the wear owing to the presence of abrasive contaminants can be reduced by the addition of polymeric powders to lubricating greases.[21]

Engine air induction filters are designed to effectively remove airborne contaminants in order to protect the engine. The engine requires a certain level of ingested air cleanliness to reduce friction and wear to improve engine efficiency. Airborne dust contaminants are very abrasive in nature and small amounts of airborne contaminants (dusts, sand etc.) can significantly increase the friction and wear in engine.[22] The efficiency of the filter has a significant effect on the wear rate and it is recommended that filters with the highest available efficiency be used. Well-designed air filters could further reduce wear and failure incidents, and hence improve the performance of internal combustion engines operating in dusty environments.[23] Lubrication is divided into three general regimes: boundary, mixed, and elastohydrodynamic/hydrodynamic.[24,25] The objective of the presented work was to study the effect of solid contaminants in engine oils on tribological parameters. The mechanism responsible for the change in friction and wear with varying particle concentration and grain size of the contaminants is discussed.

## 2. Experimental Details

### 2.1 Description of the test rig

The test rig was designed to represent the sliding reciprocating motion of the piston ring/cylinder liner interface in an engine according to standards stipulated by the ASTM G181-11.[26, 27] Figure 2 shows the designed test rig for piston ring/cylinder liner. A 1.5 kW variable speed AC motor was used to turn the crankshaft. The test specimens of the piston ring and cylinder liner were fragments from actually fired engine components, to ensure that the materials tested are the same as in the automotive engine. The piston ring was placed under the pivot arm above a reciprocating cylinder liner segment in a specially designed piston ring holder. A piezoelectric force transducer and charge amplifier were used to measure the friction force generated on the piston ring during sliding. The signals generated were received and processed using data acquisition connected to a PC to record the friction force between the ring and liner.

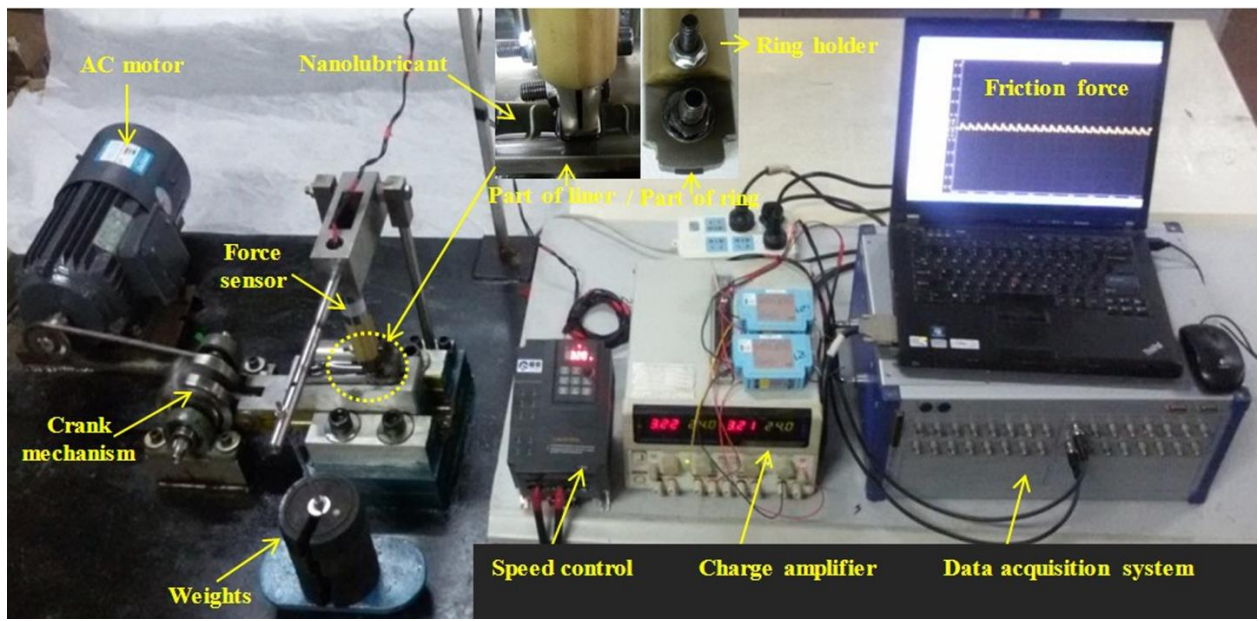

**Figure 2.** The test rig of the piston ring/cylinder liner interface designed.

The friction coefficient (μ) was evaluated as follows:

$$\mu = \frac{F_f}{F_n} \qquad (1)$$

Frictional power losses (P) were also calculated using the following formula:

$$P = F_f \cdot v_p = \mu F_n [R \omega (\sin \theta + \frac{R}{L} \sin 2\theta)] \qquad (2)$$

Where, $F_f$ is the frictional force (N), $F_n$ is the normal load applied on ring (N), $v_p$ is the reciprocating sliding speed (m s$^{-1}$), R the apparatus crank shaft radius (m), $\omega$ the angular velocity of the crank (rad s$^{-1}$), L the apparatus connecting rod length (m) and $\theta$ represents the crank angle (degrees).

## 2.2 Tribological tests characterization

An evaluation of the tribological performance of the piston ring/liner interface used the same amount of contaminated lubricant. The experimental conditions used 0.63, 0.85 and 1.1 m s$^{-1}$ average sliding speeds. In the friction tests, a normal load 120 N was used. The sliding speeds and contact loads were chosen to simulate the boundary lubrication regimes. The stroke length was fixed at 60 mm. The test specimen of piston ring and cylinder liner had a hardness of 303 HV and 257 HV respectively. The lubricating oil used for the experimental tests was a multi-grade mineral oil SAE 20W50. Oil samples contaminated with different amounts of different metals such as Al, Fe, Cu, Pb and SiO$_2$ were prepared. Concentrations of 40, 60, 80 and 100 ppm were considered with grain sizes of less than or equal to 20, 40 and 60µm. The contaminated oil samples were stirred thoroughly before being used for the friction tests. All the tests were carried out at room temperature. Acetone was used for cleaning the friction samples before testing was carried out. The cleaning was done for 20 min using a magnetic stirrer.

## 3. Results and Discussion

Figure 3 shows the tribological behavior versus crank angle with and without iron (Fe) contaminants and without combustion under 1.1 m s$^{-1}$ average sliding speed, a contaminants grain size of 20 µm, a contaminant concentration of 40 ppm and 120 N normal load.

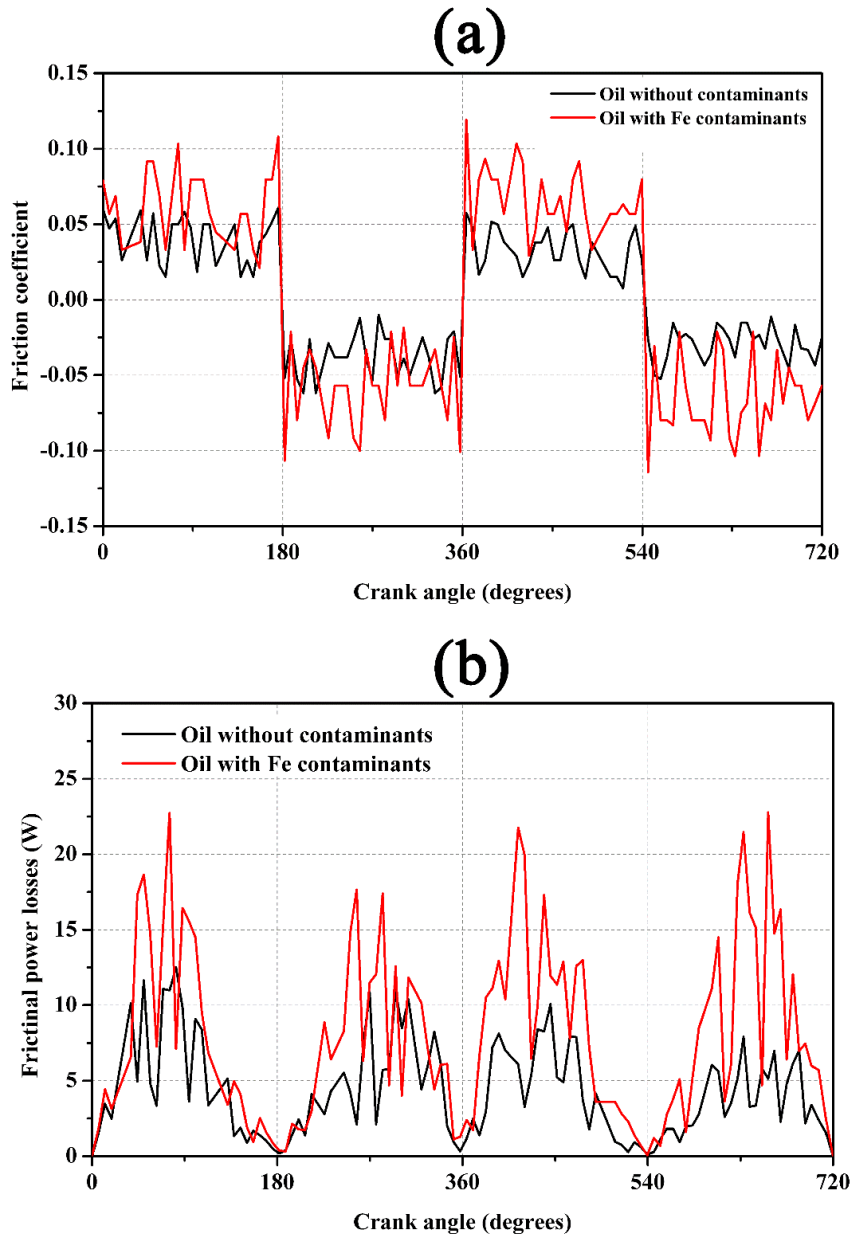

**Figure 3.** Tribological behavior versus crank angle with and without contaminants using 1.1 m s$^{-1}$ average sliding speed, contaminants grain size of 20 µm, contaminants concentration of 40 ppm and 120 N normal load.

The variations of friction coefficient with crank angle represented by the negative part of the friction curve due to the change of sliding speed direction through the reciprocating sliding motion is shown in Fig. 3 a. The maximum frictional power losses were observed at mid-stroke location, reaching zero at the top and bottom dead center (TDC and BDC) positions as shown in Fig. 3 b. This might be due to the low sliding speed of the piston. The sliding speed reached its maximum at mid-stroke causing an increase in the shear stress in the lubricant, increasing the frictional power losses as a consequence. Generally, there was an increase in the frictional power losses with an increase in engine speed. From the results it was observed that there was an increase in the boundary friction coefficient and friction power losses for oil contaminated by Fe particles because of a third body resulting in microcutting and plowing. Furthermore, an increase in the temperature of worn surface facilitates adhesive wear and plastic deformation causing the dominating flake-like and scuffing damage. The increase in friction has an adverse effect on the efficiency and fuel economy of automotive engines.

The solid contaminants were suspended in commercially available engine oil (20W50) in different concentrations (40, 60, 80 and 100 ppm). Figure 4 shows the results of the effect of solid contaminants concentrations on friction coefficient and frictional power losses using various contaminants under 0.85 m s$^{-1}$ average sliding speed and 120 N normal load. The particles used were Fe, Cu, Al, SiO$_2$, and Pb. The particle grain size was chosen to have a diameter of 20 $\mu$m. As a general trend, the average boundary friction coefficient and frictional power losses were observed to increase with an increase in the particle concentrations. It was expected that the increase of the contaminant concentration would lead to an increase in the probability of the particles to slide over each other, raising the levels of friction. Secondly, for the boundary layer thickness there would not be a separation between worn surfaces which consequently increase the level of the asperity friction. Iron particles (Fe) came first in ranking, in terms of contaminant effects, followed by copper, whilst lead came last. The reason may be related to the role of particle hardness. SiO$_2$ and Al particles came in between, and this may be attributed to the combination of rolling and plough effect of contaminant particles. Furthermore, the increase in the friction coefficient is caused by the formation of layers by the soft surface and particles, which enhances the adhesion between worn surfaces.[28]

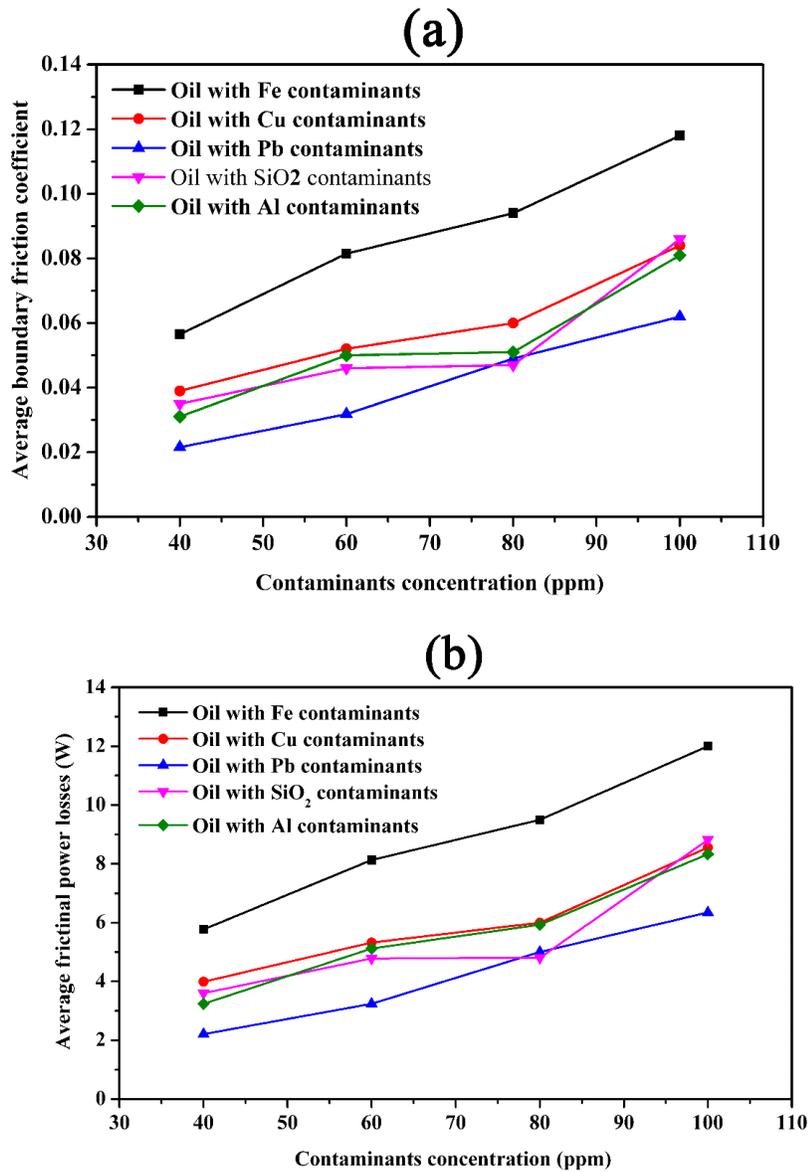

**Figure 4.** Effect of contaminants concentration on boundary friction coefficient (a) and power losses (b) under contaminants grain size of 20 µm, 0.85 m s$^{-1}$ average sliding speed and 120 N normal load.

Figure 5 indicates the relationship between the average boundary friction coefficient and contaminant grain size for different contaminant concentrations using 60 ppm. The utilized contact load and the reciprocating sliding speed were 120 N and 0.63 m s$^{-1}$ respectively. A substantial increase in the boundary friction coefficient was observed to increase with an increase in both the grain size and metal concentrations. The higher values of friction coefficient were related strongly to the iron particles. The reduction of friction coefficient values when small particles were used might receive two different explanations: one being that with small particles the indentation geometry was small (the number and size of abrasive contact); the other explanation suggested that, with small abrasive particles, clogging of the sliding system by abraded wear particles might be probable which reduced its detrimental effects.

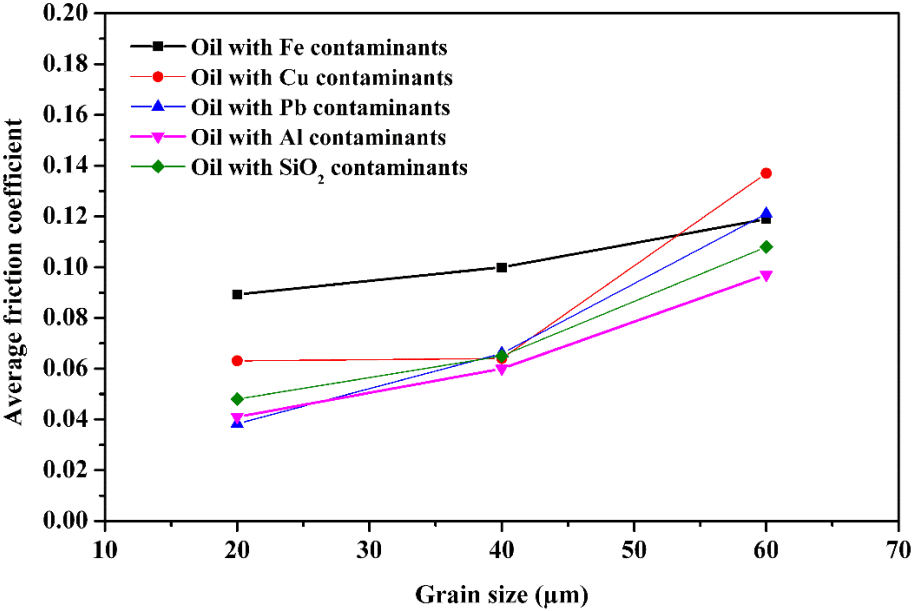

**Figure 5.** Effect of contaminants grain size on boundary friction coefficient using contaminants concentration of 60 ppm, 0.63 m s$^{-1}$ average sliding speed and 120 N normal load.

The variations of the boundary friction coefficient values for different contaminants with the average sliding velocity are illustrated in Fig. 6. The results showed that the boundary friction coefficient decreased with an increase in the average sliding speed.[29] Moreover, the results showed that Fe contaminants show friction coefficient higher.

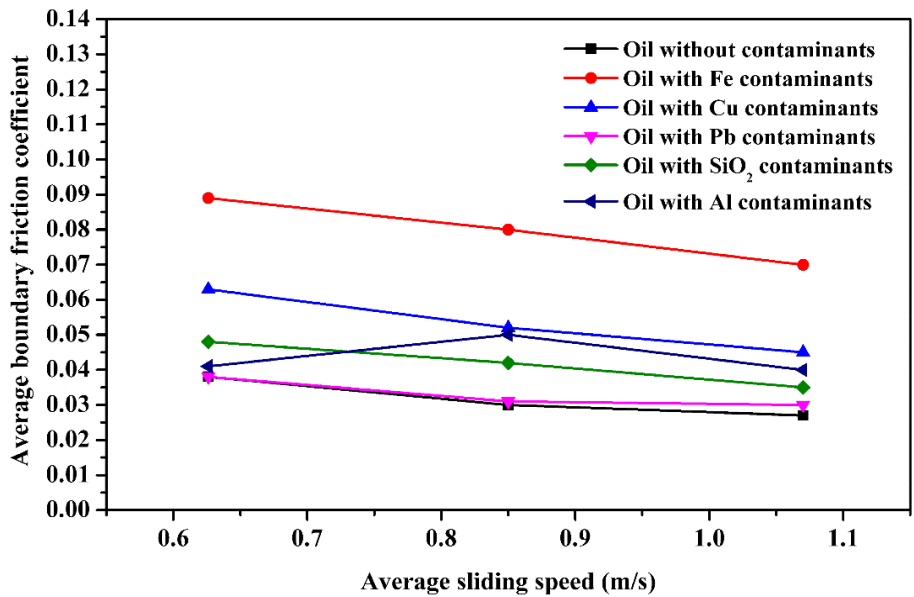

**Figure 6.** Effect of reciprocating sliding speed on average friction coefficient with different contaminants using concentration of 60 ppm, grain size of 20 µm and 120 N normal load.

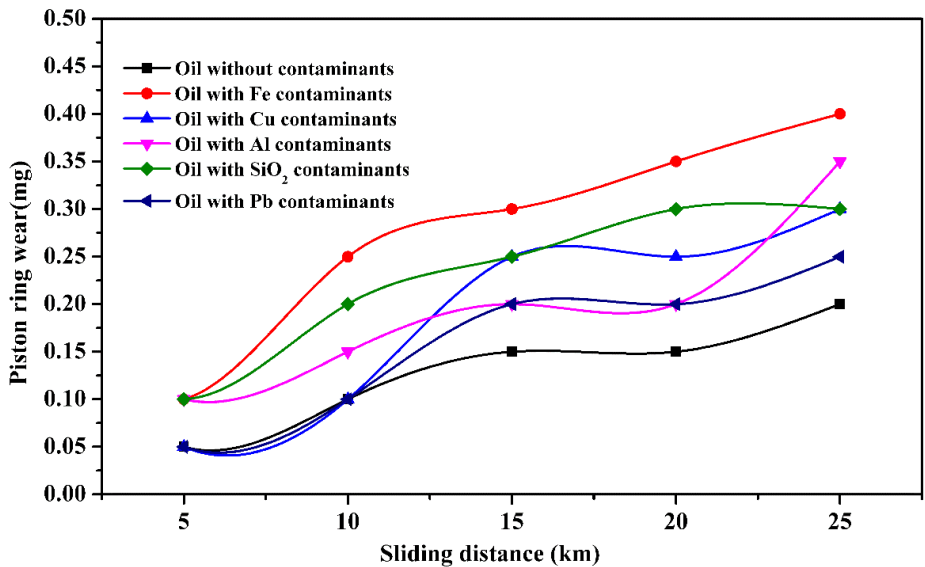

**Figure 7.** Comparisons of piston ring wear for different contaminants using concentration of 100 ppm, grain size of 20 µm and 120 N normal load.

Piston rings are an important part of most automotive engines. Commonly a set of piston rings is used to form a dynamic gas seal between the piston and cylinder liner. The sliding motion of the piston forms a thin oil film thickness between the ring land and cylinder liner, which lubricates the sliding components. With abrasive particles (metallic wear particles) entering the sliding interface between the piston ring and the cylinder liner, severe abrasive wear scars may form on the piston ring surface. The wear is normally accompanied by similar abrasive wear on the cylinder liner surface. The effect of contaminant concentrations on wear of piston ring is shown in Fig. 7 for a grain size of 20 µm and a concentration of 100 ppm. The contaminants used were Al, Fe, Cu, Pb and $SiO_2$. The contact normal load and reciprocating sliding speed were of 120 N and 0.63 m/s respectively. Generally, the wear of piston ring was observed to increase with an increase in the sliding distance. Furthermore, clean oil indicated the lowest values of ring wear whilst iron contaminants showed the highest wear levels.

## 4. Conclusions

The objective of this study was to investigate the effects of the presence of solid contaminants in lubricating oil on tribological performance during the boundary lubrication regime. A test rig was used for simulating the severe operating conditions of engine starting and stopping. Against the background of the conducted study, the following conclusions may be drawn:

**1.** The average boundary friction coefficient and frictional power losses increases with an increase in one or more of the following: contaminant concentration, grain size, and hardness of particles. The contaminant caused an increase in friction and wear because of increase of the temperature resulting from elastic and plastic deformation, repeating itself at high speed accompanied by frictional stress.

**2.** Silicon dioxide particles ($SiO_2$), although possessed a high level of hardness showed low friction behavior. The reason may be attributed to the fact that hard $SiO_2$ particles tended to roll rather than to slide.

**3.** The contaminant particles increased the abrasive wear of piston rings. The wear rate was observed to increase with an increase in grain sizes, concentrations, and sliding distance.

**4.** The filtration accuracy should be less than of the clearance between the worn surfaces to help in reducing friction and wear.